\documentclass{article}

\usepackage[final,nonatbib]{tackling_climate_workshop_style}

\usepackage[utf8]{inputenc} 
\usepackage[T1]{fontenc}    
\usepackage{hyperref}       
\usepackage{url}            
\usepackage{booktabs}       
\usepackage{amsfonts}       
\usepackage{nicefrac}       
\usepackage{microtype}      
\usepackage{algorithm}
\usepackage{algpseudocode}
\usepackage{siunitx}
\usepackage{amsmath}
\usepackage{bm}

\usepackage[pdftex]{graphicx}

\title{Hyperspectral shadow removal with Iterative Logistic Regression and latent Parametric Linear Combination of Gaussians}

\author{
  Core Francisco Park\\
  Department of Physics\\
  Harvard University\\
  Cambridge, MA 02138, U.S.A.\\
  \texttt{corefranciscopark@g.harvard.edu}\\
  \And
  Maya Nasr\\
  John A. Paulson School of\\
  Engineering and Applied Sciences\\
  Harvard University\\
  Cambridge, MA 02138, U.S.A.\\
  \texttt{mayanasr@g.harvard.edu}\\
  \And
  Manuel Pérez-Carrasco\\
  Data Science Unit\\
  University of Concepcion\\
  Biobio, Chile\\
  maperezc@inf.udec.cl\\
  \And
  Eleanor Walker\\
  John A. Paulson School of\\
  Engineering and Applied Sciences\\
  Harvard University\\
  Cambridge, MA 02138, U.S.A.\\
  \texttt{ewalker@g.harvard.edu}\\
  \And
  Douglas Finkbeiner\\
  Department of Physics\\
  Harvard University\\
  Cambridge, MA 02138, U.S.A.\\
  \texttt{dfinkbeiner@cfa.harvard.edu}\\
  \And
  Cecilia Garraffo\\
  AstroAI\\
  Center for Astrophysics $|$ Harvard \& Smithsonian\\
  Cambridge, MA 02138, U.S.A.\\
  \texttt{cgarraffo@cfa.harvard.edu}\\
}

\begin{document}

\maketitle

\begin{abstract}
Shadow detection and removal is a challenging problem in the analysis of hyperspectral images. Yet, this step is crucial for analyzing data for remote sensing applications like methane detection. In this work, we develop a shadow detection and removal method only based on the spectrum of each pixel and the overall distribution of spectral values. We first introduce Iterative Logistic Regression (ILR) to learn a spectral basis in which shadows can be linearly classified. We then model the joint distribution of the mean radiance and the projection coefficients of the spectra onto the above basis as a parametric linear combination of Gaussians. We can then extract the maximum likelihood mixing parameter of the Gaussians to estimate the shadow coverage and to correct the shadowed spectra. Our correction scheme reduces correction artefacts at shadow borders. The shadow detection and removal method is applied to hyperspectral images from MethaneAIR, a precursor to the satellite MethaneSAT.
\end{abstract}

\section{Introduction}
Methane is the second most important greenhouse gas following carbon dioxide. MethaneSAT is a satellite mission aiming to map the methane density in the atmosphere\cite{rohrschneider2021methanesat}. For this and other similar missions, careful analysis of the hyperspectral image resulting from the detector is essential to retrieve the target atmospheric components. Two significant challenges in image analysis are clouds and shadows\cite{chulakadabba2023methane}. In the case of MethaneSAT, a cloud essentially blocks the retrieval of methane density under it. However, depending on the indirect ground illumination, shadowed pixels might carry a detectable signal. These pixels could either be processed by a different downstream pipeline or be corrected for the effect of the shadow before entering the subsequent pipelines. In this work, we present a probabilistic latent spectral space inference method to remove the average effect of a shadow on a spectrum, while preserving the original spectral features as much as possible. This method is applied to hyperspectral images from MethaneAIR, a precursor to the MethaneSAT satellite.

\section{Method}
\begin{figure}
  \centering
  \includegraphics[width=\linewidth]{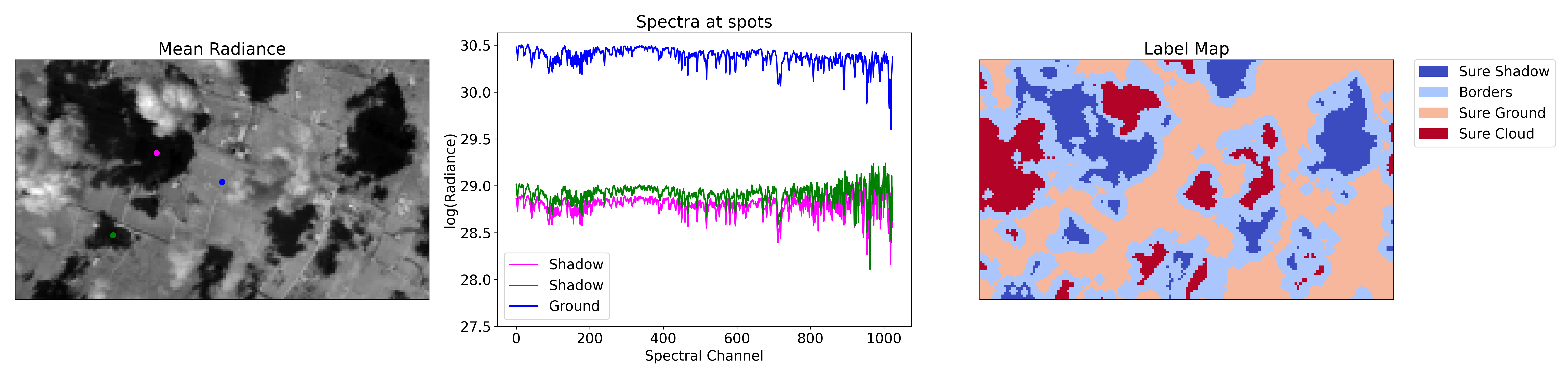}
  \caption{Hyperspectral data from MethaneAIR. (Left) The mean radiance for all pixels. (Middle) The spectrum at each of the 3 points on the left image. (Right) The map of labels. ``Sure Ground" and ``Sure Cloud" are simple made by binary erosion of the shadow and ground labels.}
  \label{fig:fig1}
\end{figure}

\paragraph{Data}
We use hyperspectral imaging data from a MethaneAIR flight \cite{chulakadabba2023methane}. Our data is a set of hyperspectral images with 1024 spectral bins and varying width and height. An image with clouds and shadows within the field of view is in Fig.~\ref{fig:fig1}. The most evident feature of a shadow is the reduced mean radiance. To enhance the method to learn spectral shape changes by shadows, we explicitly remove the mean radiance ($m_i$) by dividing each spectrum by its mean and taking the logarithm. Thus each spectrum $\bm{f}_i$ is normalized as $\bm{s}_i=\log(\bm{f}_i/m_i)$. $\log$ is the natural logarithm everywhere. 
\paragraph{Iterative Logistic Regression and Latent Basis Learning}
Using spectra and binary labels($\bm{s}_i\in\mathcal{R}^{1024},\: y_{i} \in [0,1]$) from sunlit and shadowed pixels, we train our Iterative Logistic Regression(ILR), described in Algorithm \ref{alg:ILR}. On each iteration, the best spectral weight separating shadowed pixels and bright pixels are learned by a logistic regression and projected out from the data, until the classification accuracy becomes negligible. The resulting weights($\bm{W}\in\mathcal{R}^{(k,1024)}$) form a k-dimensional spectral basis for the latent space and the projection coefficients of each spectrum onto this basis($\bm{\beta}_i\in\mathcal{R}^k\:s.t.\:\bm{s}_i=\bm{\beta}_i^T \bm{W}+\bm{s}^{null}_i$) becomes the features of each spectrum in this basis. In practice the dimensionality of this space is much smaller than the number of spectral bins, in our case $k=23$.
\begin{algorithm}
\caption{Iterative Logistic Regression}\label{alg:ILR}
\begin{algorithmic}
\Require $x\in \mathcal{R}^{(n,p)},\:y\in [0,1]^{n},\:f_{thres}\in(0,1)$ \Comment{Dataset with n rows and p features}
\State $W=[];\:f_1=1$
\While{$f_1 \geq f_{thres}$}\Comment{Repeat until the data is hard to classify}
    \State $x_{Tr},y_{Tr},x_{Te},y_{Te}=split(x,y)$
    \State $w=get\_w_LR(x_{Tr},y_{Tr})$\Comment{Get the separating hyperplane from logistic regression}
    \State $\hat{y_{Te}}=\sigma (x_{Te}\cdot w)>0.5;\: f_1=get\_f_1(y_{Te},\hat{y}_{Te})$ \Comment{Get the predictions, $\sigma$ is a sigmoid}
    \State $x\leftarrow x-x\cdot w/||w||_2$\Comment{Project out this component}
    \State $W.append(w)$
\EndWhile
\State \Return matrix(W)
\end{algorithmic}
\end{algorithm}
\paragraph{Joint estimation of shadow fraction and mean radiance for border pixels.}
Now, we construct a new feature space for each pixel by combining the log average radiance ($\log(m_i)$) and $\bm{\beta}_i$. This is our (k+1) dimensional latent spectrum, $\bm{e}_i\in\mathcal{R}^{k+1}$, with $e_{i,0}=\log(m_i)$. We approximate the joint distribution for the sunlit latent spectrum (subscript g for ground) and the shadowed latent spectrum (subscript s for shadow) as Gaussian distributions. Our model then assumes each latent spectrum on the border is a draw from a linear combination of these Gaussians. It is important to note that our model is not a Gaussian Mixture Model(GMM) but simply a Parametric Linear Combination of Gaussians(PLCG).
\begin{align}
    \bm{e}^b_i&\sim (1-\alpha_i) \mathcal{N}_g(\bm{\mu}_g,\bm{\Sigma}_g)+\alpha_i\mathcal{N}_s(\bm{\mu}_s,\bm{\Sigma}_s)\quad\quad(0\leq \alpha \leq 1)
\end{align}
where $\bm{e}^b_i$ is a border latent spectrum, $\alpha$ is the mixing parameter and $\bm{\mu}_g,\bm{\Sigma}_g,\bm{\mu}_s,\bm{\Sigma}_s$ are the means and covariances for the ground and shadowed latent spectrum, respectively. The superscript $i$ indicates that each border latent spectrum is assumed to be drawn from a distribution resulting from a different mixing parameter. When $\alpha=0$ the border latent spectrum is just a draw from $\mathcal{N}_g$ and for $\alpha=1$ it is a draw from $\mathcal{N}_s$, thus $
\alpha$ is a parameter representing how much a spectrum is covered by a shadow. We can then estimate $\alpha$ independently for each border latent spectrum using maximum likelihood estimation (MLE) under a Gaussian likelihood\footnote{This assumes the two distributions are jointly Gaussian distributed}.
\begin{align}
    \alpha^{*}_i&=argmax_{\alpha_i}\:\mathcal{L}(\alpha_i|\bm{e}^b_i)\\
    \log\mathcal{L}(\alpha_i|\bm{e}^b_i)&=-\frac{1}{2}\log(|\bm{\Sigma}(\alpha)|)-\frac{1}{2}(\bm{e}^b_i-\bm{\mu}(\alpha))^T\bm{\Sigma}^{-1}(\alpha)(\bm{e}^b_i-\bm{\mu}(\alpha))+C\label{eq:likelihood}\\
    \bm{\mu}(\alpha)&=(1-\alpha)\bm{\mu}_g+\alpha\bm{\mu}_s\quad\quad\bm{\Sigma}(\alpha)=(1-\alpha)\bm{\Sigma}_g+\alpha\bm{\Sigma}_s
\end{align}
Since the latent space's dimension is small($\mathcal{O}(10)$), and $\alpha$ is a single parameter, a simple grid search can be performed to find the (MLE) $\alpha$. Then each corrected border spectrum (and its mean) can be obtained as:
\begin{align}
    \bm{e'}^b_i&=\bm{S}_i(\bm{e}^b_i-\bm{\mu}(\alpha^{*}_i))+\bm{\mu}_g\\
    \bm{s'}^b_i&=(\bm{e'}^b_{i,1:k})^T W+\bm{s}_i^{null}\quad \mathrm{and}\quad \log(\Bar{s}_i)={e'}^b_{i,0}
\end{align}
where $\bm{S}$ is an identity matrix everywhere except $S_{00}=\sqrt{\Sigma_{g00}/\Sigma(\alpha^{*}_i)_{00}}$. This $S_{00}$ term scales $e_{i,0}$ by the ratio of standard deviation of $\log(m_i)$ of ground spectra to shadow spectra, so that the whole region under a shadow doesn't end up with a lower contrast. This is motivated by the fact that we only want to correct for the average effect of the shadows while keeping the spectral variances as they are.

\section{Results}

\begin{figure}[h]
  \centering
  \includegraphics[width=\linewidth]{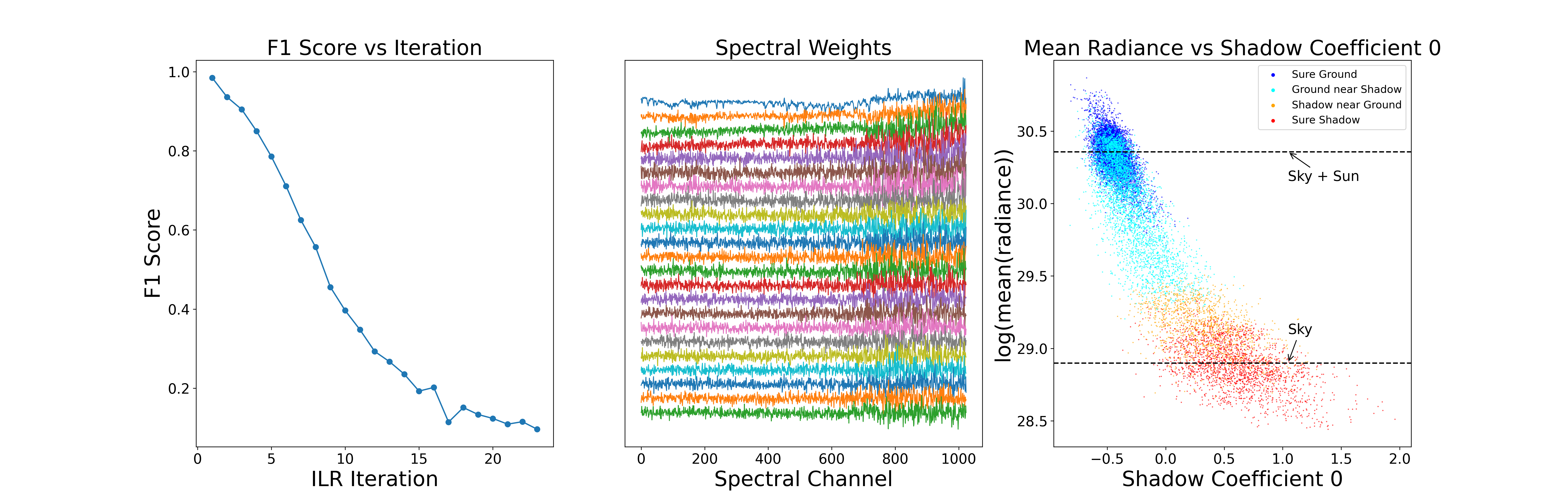}
  \caption{Iterative Logistic Regression and the shadow basis (Left) F1-score as components relevant to shadow classification are iteratively removed. (Middle) The learned spectral weights($\bm{W}$) forming the shadow basis. (Right) The joint distribution of the log-mean radiance and the first shadow coefficient($(m_i,e^b_{i,1})$). We observe that the sunlit spectra is roughly $e^{1.5}\sim4.5$ times brighter than the shadowed spectra.}
  \label{fig:fig2}
\end{figure}

\begin{figure}[h]
  \centering
  \includegraphics[width=0.9\linewidth]{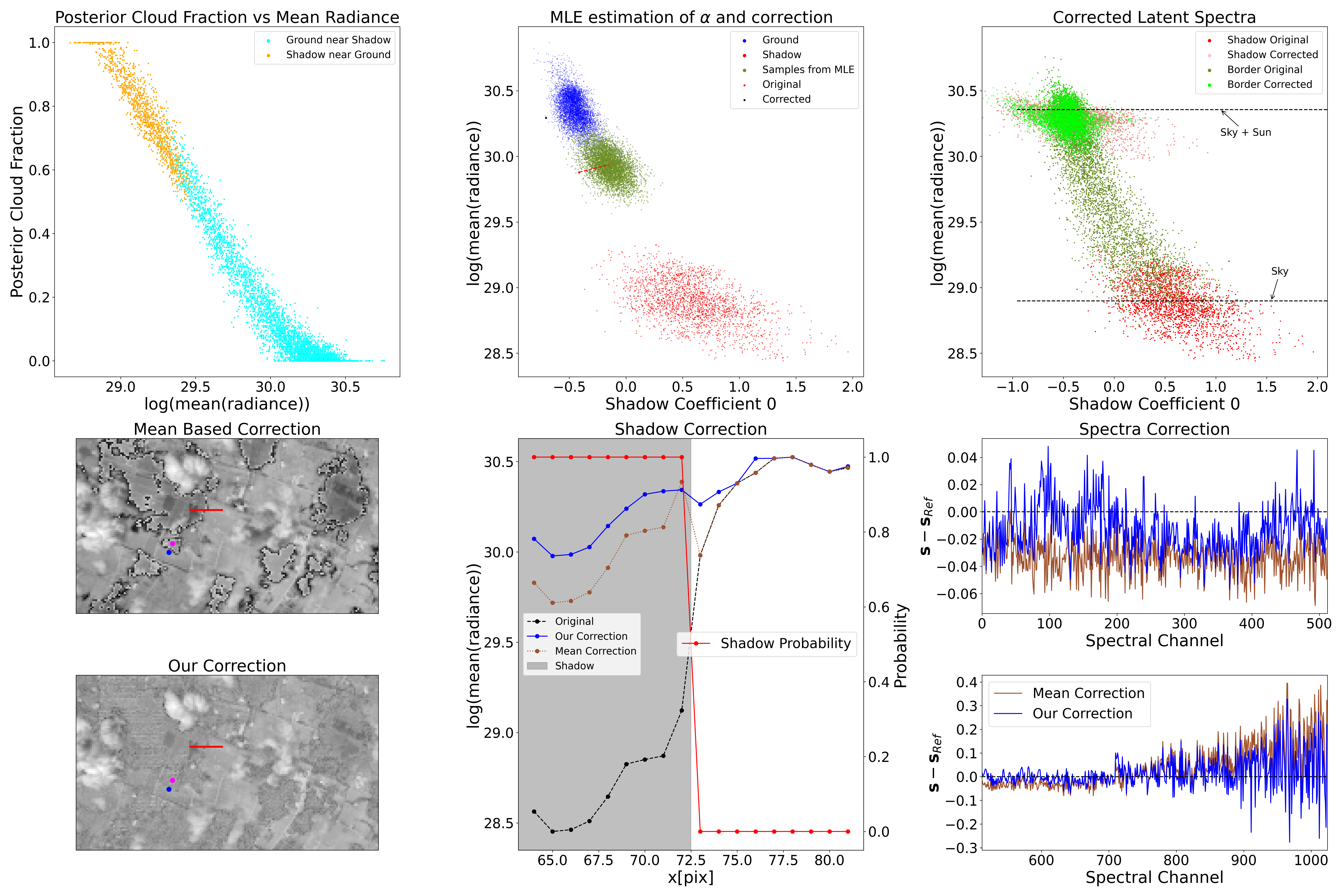}
  \caption{Our shadow correction method. (Top Left) The inferred posterior cloud fraction as a function of mean radiance. (Top Center) Our correction method illustrated. The olive colored points are samples from the inferred mixture of Gaussians. (Top Right) The distribution shift in latent space from our correction method. (Bottom Left) the mean based correction and our correction (Bottom Center) Our correction method across a shadow border(red line). (Bottom Right) Spectral correction from our method. The spectra are corrected from the magenta point and is compared with the reference spectrum from the blue point.}
  \label{fig:fig3}
\end{figure}

\paragraph{ILR and shadow basis learning}
Fig.~\ref{fig:fig2} shows ILR applied to MethaneAIR data with ground and shadow labels. After $k=23$ iterations, the F1-score is negligible and we have extracted the basis vectors relevant for linear shadow classification. From the joint distribution of the mean radiance and the shadow coefficients of every pixel, we find that the ``sure ground" and ``sure shadow" can both be described as a (k+1)-dimensional Gaussian distribution, while the border pixels are scattered between them.
\paragraph{Shadow Correction}
Fig.~\ref{fig:fig3} illustrates our correction method and its results. We find that the posterior cloud fraction is, as expected, a tight function of the mean radiance. We find that our reconstruction scheme is better than adjusting the mean radiance by a factor determined from the raw probabilities of logistic regression. This is illustrated in the bottom center panel of Fig.~\ref{fig:fig3}, where the cloud probability is discontinuous and does contain information about the shift in the mean radiance. A mean correction thus results in an over-correction right inside the shadow and a under-correction right outside the shadow. Our method mitigates this effect.

\section{Discussion}
\paragraph{Objective of shadow removal}
Our method casts shadow removal into a slightly different problem than in the literature. Previous works \cite{zhang_2020_shadow,zhao_2021_shadow,duan_2022_shadow}, focus on retrieving the radiance and spectral shape which would have been measured if the shadow was not present. Here, we focus on removing the average effect of having a shadow cast on a pixel, while conserving the original spectrum as much as possible.

\paragraph{Impact towards tackling climate change}
Our approach increases the raw amount of area from which methane emission can be detected\cite{chulakadabba2023methane}. MethaneAIR has two spectrometers targeting 1236-1319 nm for $O_2$ and 1592-1697 nm for $CO_2$ and $CH_4$ (1249-1305 nm and 1598-1683 nm respectively for MethaneSAT). MethaneAIR detects the total column-averaged dry-air mole fractions of $CH_4$ ($XCH_4$) by the $CO_2$ proxy method \cite{frankenberg2005iterative} using the second (longer wavelength) spectrometer. This requires accurate spectra at the P and R bands of $CO_2$ (1595–1618 nm) and the $2\nu_3$ band of $CH_4$ (1629–1654 nm) \cite{chan2023methane}. These bands correspond roughly to spectral channel indices 50-250 and 350-550 in the x-axis of Fig.~\ref{fig:fig3}, bottom right. We find that our correction method is significantly different from the mean based correction in these regions and thus demonstrates that it might affect the methane detection significantly. We plan to study this more carefully in a future work.

\paragraph{Limitations and Future Work}
A major limitation of the presented model is that it is agnostic about any spatial priors, and this limits the estimation of the shadow coverage at the borders. This is the main reason our shadow removal, although better than merely correcting the mean of the distribution, has spatial artifacts at the shadow borders. We believe the introduction of spatial priors added to the likelihood function in Eq.~\ref{eq:likelihood} will improve our method and further reduce over/under-corrections.

Our method doesn't make use of any assumptions specific to shadows, so the same model can be used to correct for other continuous artifacts.

\section{Conclusion}
We have developed a method to detect and remove the average effect of shadows from a hyperspectral image using only binary shadow and ground labels. We plan to enhance the model using spatial priors and apply it to MethaneAIR and MethaneSAT images for additional detection of methane in shadowed areas.

\begin{ack}
CFP gratefully acknowledges Dr. Steven Wofsy and Christopher Chan Miller for providing the data used in this work and for useful discussions. This work is supported by the Environmental Defense Fund and by MethaneSAT.org, a wholly-owned subsidiary of the Environmental Defense Fund. The computations in this paper were run on the FASRC Cannon cluster supported by the FAS Division of Science Research Computing Group at Harvard University. This collaboration was facilitated by EarthAI and AstroAI.
\end{ack}

\bibliographystyle{unsrt}
\bibliography{references}

\end{document}